\documentclass[10pt]{article}
\usepackage{amsmath}
\usepackage{cite}
\usepackage{relsize}
\usepackage{amssymb}
\usepackage{graphics}
\usepackage{epsfig}
\usepackage{epstopdf}
\usepackage{verbatim}
\usepackage{color}
\usepackage{subcaption}
\usepackage{multirow}
\usepackage{textcomp}
\usepackage{float}
\usepackage{mathtools}
\usepackage[toc,page]{appendix}
\usepackage{enumitem}

\begin{document}
\title{Highly predictive and testable $A_{4}$ flavor model within type-I and II seesaw framework and associated phenomenology}
\author {Surender Verma\thanks{Electronic address: s\_7verma@yahoo.co.in}, Monal Kashav\thanks{Electronic address: monalkashav@gmail.com}  and Shankita Bhardwaj\thanks{Electronic address: shankita.bhardwaj982@gmail.com}}

\date{\textit{Department of Physics and Astronomical Science,\\Central University of Himachal Pradesh, Dharamshala 176215, INDIA.}}
\maketitle

\begin{abstract}
We investigate neutrino mass model based on $A_4$ discrete flavor symmetry in type-I+II seesaw framework. The model has imperative predictions for neutrino masses, mixing and $CP$ violation testable in the current and upcoming neutrino oscillation experiments. The important predictions of the model are: normal hierarchy for neutrino masses, a higher octant for atmospheric angle
($\theta_{23}>45^{o}$) and near-maximal Dirac-type $CP$ phase ($\delta\approx\pi/2$ or $3\pi/2$) at $3\sigma$ C. L.. These predictions are in consonance with the latest global-fit and results from Super-Kamiokande(SK), NO$\nu$A and T2K. Also, one of the important feature of the model is the existence of a lower bound on effective Majorana mass, $|M_{ee}|\geq 0.047$eV(at 3$\sigma$) which corresponds to the lower part of the degenerate spectrum and is within the sensitivity reach of the neutrinoless double beta decay(0$\nu\beta\beta$) experiments. 
\end{abstract}
\maketitle
\section{Introduction}
\par The neutrino oscillation experiments have conclusively demonstrated that neutrinos have tiny mass and they do mix. Especially, with the observation of non-zero $\theta_{13}$\cite{t1, t2, t3, t4, t5, t6} the $CP$ conserving part of the neutrino mixing matrix is known to high precision: $\theta_{12}=34.5^{+1.2}_{-1.0}$, $\theta_{23}= 47.7^{+1.2}_{-1.7}$, $\theta_{13}=8.45^{+0.16}_{-0.14}$\cite{data}. Although, the two neutrino mass squared differences $\Delta m^{2}_{12}$ and $|\Delta m^{2}_{23}|$ have, also, been measured but there still exist two possibilities for neutrino masses to be either normal hierarchical(NH) or inverted hierarchical(IH).

\par Understanding this emerged picture of neutrino masses and mixing, which is at odds with that characterizing the quark sector, is one of the biggest challenge in elementary particle physics. The Yukawa couplings are undetermined in the gauge theories. To understand the origin of neutrino mass and mixing one way is to employ phenomenological approaches such as texture zeros\cite{t10,t20,t30,t40,t50,t60,t70,t80,t90,t100,t110,t120,t130,t140,t150,t160}, hybrid textures \cite{ht1,ht2,ht3,ht4,ht5}, scaling\cite{sc1,sc2,sc3,sc4,sc5,sc6,sc7,sc8}, vanishing minor\cite{v1,v2,v3} etc. irrespective of details of the underlying theory. These different ansatze are quite predictive as they decrease the number of free parameters in neutrino mass matrix. The second way, which is more theoretically motivated, is to apply yet-to-be-determined non-Abelian flavor symmetry. In this approach a flavor symmetry group is employed in addition to the gauge group to restrict the Yukawa structure culminating in definitive predictions for values and/or correlations amongst low energy neutrino mixing parameters.

\par Discrete symmetry groups have been successfully employed to explain non-zero tiny neutrino masses and large mixing angle in lepton sector\cite{d,d1,d2,d3,d4,d5,d6,d7,d8}. There exist plethora of choices for flavor groups having similar predictions for neutrino masses and mixing patterns. In general, a flavor model results in proliferation of the Higgs sector making it sometime discouragingly complex. The group $A_4$\cite{a41,a42,a43,a44} being the smallest group having 3-dimensional representation is widely employed as the possible underlying symmetry to understand neutrino masses and mixing with in the paradigm of seesaw mechanism\cite{seesaw1a,seesaw1b,seesaw1c,seesaw1d,seesaw2a,seesaw2b,seesaw2c,seesaw2d,seesaw2e}. It has been successfully employed to have texture zero(s) in the neutrino mass matrix which is found to be very predictive\cite{t80,t130,t140,t150}.

\par Another predictive ansatz is hybrid texture structure with one equality amongst elements and one texture zero in neutrino mass matrix. The hybrid texture of the neutrino mass matrix has been realized under $S_3\otimes Z_3$ symmetry with in type-II seesaw framework assuming five scalar triplets with different charge assignments under $S_3$ and $Z_3$\cite{ht4}. Also, some of these hybrid textures have been realized under Quaternion family symmetry $Q_8$\cite{Q8}. In particular, the authors of Ref. \cite{ht4} realized one of such hybrid texture with non-minimal extension in the scalar sector of the model which requires the imposition of an additional cyclic symmetry to write group-invariant Lagrangian. Also, the vacuum alignments have not been shown to be realizable. Keeping in view existing gaps, we are encouraged for realization of hybrid textures with minimal extension of scalar sector under group $A_4$(smallest group with 3-dimensional representation) assuming the vacuum alignments $\langle \Phi_{0} \rangle=\frac{\vartheta}{\sqrt{3}} (1,1,1)^{T}$. In fact, for $A_4$ flavor model, it has been shown in Ref. \cite{vev1} that this VEV minimizes the scalar potential. In this work, for the first time, we have employed $A_4$ flavor symmetry to realize hybrid texture structure of neutrino mass matrix. We present a simple minimal model based on group $A_4$ with two right-handed neutrinos in type-I+II seesaw mechanism leading to hybrid texture structure for neutrino mass matrix. The same Higgs doublet is responsible for the masses of charged leptons and neutrinos\cite{t80}. In addition, one scalar singlet Higgs field $\chi$ and two scalar triplets $\Delta_i$($i=1,2$) are required to write $A_4$ invariant Lagrangian. 

 \par In Sec. II, we systematically discuss the model based on group $A_4$ and resulting effective Majorana neutrino mass matrix. Sec. III is devoted to study phenomenological consequences of the model. In this section we, also, study the implication to neutrinoless double beta decay (0$\nu\beta\beta$) process. Finally, in Sec. IV, we summarize the predictions of the model and their testability in current and upcoming neutrino oscillation/$0\nu\beta\beta$ experiments.
 
 \section{The $A_4$ Model}
The group $A_{4}$ is a non-Abelian discrete group of even permutations of four objects. It has four conjugacy classes, thus, have four irreducible representations(IRs), viz.: $\bf{1}$, $\bf{1'}$, $\bf{1''}$ and $\bf{3}$. The multiplication rules of the IRs are: \textbf{1$'$$\otimes$1$'$ =1$''$}, \textbf{1$''$$\otimes$1$''$=1$'$}, \textbf{1$'$$\otimes$1$''$=1}, 
\textbf{3$\otimes$3=1$\oplus$1$'$$\oplus$1$''$$\oplus$$\bf{3}_s$$\oplus$$\bf{3}_a$} where,  
\begin{eqnarray}
\nonumber
&&\left(\bf{3}\otimes\bf{3}\right)_{\bf{1}}=a_1b_1+a_2b_2+a_3b_3,\\ \nonumber
&&\left(\bf{3}\otimes\bf{3}\right)_{\bf{1'}}=a_1b_1+\omega a_2b_2+\omega^2 a_3b_3,\\ \nonumber
&&\left(\bf{3}\otimes\bf{3}\right)_{\bf{1''}}=a_1b_1+\omega^2 a_2b_2+\omega a_3b_3,\\ \nonumber
&&\left(\bf{3}\otimes\bf{3}\right)_{\bf{3_s}}=\left(a_2b_3+b_2a_3,a_3b_1+a_1b_3,a_1b_2+a_2b_1\right),\\ \nonumber
&&\left(\bf{3}\otimes\bf{3}\right)_{\bf{3_a}}=\left(a_2b_3-b_2a_3,a_3b_1-a_1b_3,a_1b_2-a_2b_1\right),\\ \nonumber
\end{eqnarray}
$\omega\equiv e^{2\pi i/3}$ and $\left(a_1,a_2,a_3\right)$, $\left(b_1,b_2,b_3\right)$ are basis vectors of the two triplets. Here, we present an $A_{4}$ model within type-I+II seesaw framework of neutrino mass generation. In this model, we employed one $SU(2)_{L}$ 
Higgs doublet $\Phi$, one $SU(2)_L$ singlet Higgs $\chi$ and two $SU(2)_{L}$ triplet Higgs fields($\Delta_{1},\Delta_{2}$). The transformation properties of different fields under $SU(2)_{L}$ and $A_4$ are given in Table \ref{tab1}. These field assignments under $SU(2)_L$ and $A_4$ leads to the following Yukawa Lagrangian 

\begin{table}
\centering
\begin{tabular}{ccccccccccc}
 Symmetry & $D_{iL}$ & $e_{R}$ & $\mu_{R}$ & $\tau_{R}$ & $\nu_{1}$ & $\nu_{2}$ & $\chi$ &$\Phi$& $\Delta_{1}$ & $\Delta_{2}$\\ \hline
$SU(2)_L$    &     2      &    1      &   1     &    1        &   1        &    1   & 1 &      2   & 3           & 3        \\ \hline
$A_{4}$ &      3     &      1    &     1$'$      &     1$''$       &       1       &       1$''$ &       1$''$   &     3    &    1     &   1$''$     
\end{tabular}
\caption{\label{tab1} Field content of the model and charge assignments under $SU(2)_L$ and $A_{4}$.} 
\end{table}
\begin{eqnarray}
\nonumber
 - \mathcal{L}= &&y_{e}(\bar{D}_{eL} \phi_{1}+\bar{D}_{\mu L} \phi_{2}+\bar{D}_{\tau L} \phi_{3})_{\bf{1}} e_{R_{\bf{1}}} \\
 \nonumber
&&+y_{\mu}(\bar{D}_{eL} \phi_{1}+\omega^{2} \bar{D}_{\mu L} \phi_{2}+ \omega \bar{D}_{\tau L} \phi_{3})_{\bf{1''}} \mu_{R_{\bf{1'}}} \\
\nonumber
 &&+y_{\tau}(\bar{D}_{eL} \phi_{1}+\omega \bar{D}_{\mu L} \phi_{2}+ \omega^{2} \bar{D}_{\tau L} \phi_{3})_{\bf{1'}} \tau_{R_{\bf{1''}}} \\
 \nonumber
 && +y_{1} (\bar{D}_{eL} \tilde{\phi}_{1}+\bar{D}_{\mu L} \tilde{\phi}_{2}+\bar{D}_{\tau L} \tilde{\phi}_{3})_{\bf{1}} \nu_{1_{\bf{1}}} \\
 \nonumber
 && +y_{2} (\bar{D}_{eL} \tilde{\phi}_{1}+\omega \bar{D}_{\mu L} \tilde{\phi}_{2}+ \omega^{2} \bar{D}_{\tau L} \tilde{\phi}_{3})_{\bf{1'}} \nu_{2_{\bf{1''}}} \\
 \nonumber
 && -y_{\Delta_{1}} ({D}_{eL}^T  C^{-1} {D}_{eL} + {D}_{\mu L}^{T} C^{-1} {D}_{\mu L} +{D}_{\tau L}^T C^{-1} {D}_{\tau L})_{\bf{1}}i \tau_{2} \Delta_{1_{\bf{1}}}\\
 \nonumber
 && -y_{\Delta_{2}} ({D}_{eL}^{T} C^{-1} {D}_{eL} + \omega {D}_{\mu L}^{T} C^{-1} {D}_{\mu L}+ \omega^{2} {D}_{\tau L}^{T}C^{-1} {D}_{\tau L})_{\bf{1'}}i \tau_{2} \Delta_{2_{\bf{1''}}} \\
 && - M {\nu}_{1}^{T} C^{-1} {\nu}_{1}-h_\chi \chi {\nu}_{2}^{T} C^{-1} \nu_{2}+h.c.\\
\nonumber
 \end{eqnarray}
where, $\tilde{\phi}=i\tau_2\phi^*$ and $y_{i} (i= e,\mu,\tau,1,2,\Delta_{1},\Delta_{2}$) are Yukawa coupling constants.

\par The above Lagrangian leads to charged lepton mass matrix $m_{l}$, right handed Majorana 
mass matrix $m_{R}$ and Dirac mass matrix $m_{D}$ given by\\
\begin{equation}
 m_{l}= U_{L} diag(y_{e},y_{\mu},y_{\tau})\vartheta,
\end{equation}
 \begin{equation}
 m_{R}=
  \begin{pmatrix}
    M  & 0  \\
    0 & N \\
   \end{pmatrix},
\end{equation}
\begin{equation}
 m_{D}={\begin{pmatrix}
    x   & y \\
    x  & \omega y \\
    x & \omega^{2} y  
   \end{pmatrix}},
\end{equation}
after spontaneous symmetry breaking with vacuum expectation values(VEVs) as $\langle \Phi_{0} \rangle=\frac{\vartheta}{\sqrt{3}} (1,1,1)^{T} $ and $\langle \chi_0\rangle=\varepsilon$ for Higgs doublet and scalar singlet, respectively. It has been thoroughly studied in literature \cite{vev1,vev2,vev3} that  vacuum expectation value  $\frac{\vartheta}{\sqrt{3}} (1,1,1)^{T} $  minimizes $A_4$ scalar potential. Here $U_{L}$ is 
\begin{equation}
\frac{1}{\sqrt{3}}
  \begin{pmatrix}
    1  & 1 & 1  \\
    1 & \omega^{2} & \omega   \\
    1 & \omega & \omega^{2}
   \end{pmatrix},
\end{equation}
which diagonalizes $m_l$, $N=h_{\chi} \varepsilon$, $x=\frac{\vartheta}{\sqrt{3}} y_{1}$ and $y= \frac{\vartheta}{\sqrt{3}} y_{2}$.
The type-I seesaw contribution to effective Majorana neutrino mass matrix is 
\begin{equation}
 m_{\nu_{1}} = m_{D} m_{R}^{-1} m_{D}^{T}.
\end{equation}
Using Eqns.(3) and (4) we get
\begin{equation}
m_{\nu_{1}}= {
  \begin{pmatrix}
    \frac {x^{2}}{M}+\frac{y^{2}} {N} & \frac{x^{2}}{M}+ \frac{\omega y^{2}}{N}  & \frac{x^{2}}{M}+ \frac{\omega^{2} y^{2}}{N}  \\
   \frac{ x^{2}}{M}+\frac{\omega y^{2}}{N} & \frac{x^{2}}{M}+\frac{\omega^{2} y^{2}}{N} & \frac{x^{2}}{M}+\frac{y^{2}}{N}  \\
    \frac{x^{2}}{M}+ \frac{\omega^{2} y^{2}}{N} & \frac{x^{2}}{M}+\frac{y^{2}}{N} & \frac{x^{2}}{N}+\frac{\omega y^{2}}{N}
   \end{pmatrix}}.
\end{equation}
Assuming VEVs $\upsilon_{j}$($j=1,2$) for scalar triplets $\Delta_{1}, \Delta_{2}$, respectively, the type-II seesaw contribution to effective Majorana mass matrix is 
\begin{equation}
m_{\nu_{2}}=
  \begin{pmatrix}
    c+d  & 0 & 0  \\
    0 & c+\omega d &  \\
    0 & 0& c+\omega^{2} d
   \end{pmatrix},
\end{equation}
where $c=y_{\Delta_{1}}\upsilon_{1}$ and $d= y_{\Delta_{2}}\upsilon_{2}$.
So, effective Majorana mass matrix is given as
$$ m_{\nu}=m_{\nu_{1}}+m_{\nu_{2}} .$$
The charge lepton mass matrix $m_{l}$ can be diagonalized by the transformation 
$$ M_{l}= U_{L}^{\dagger} m_{l}U_{R},$$
where $U_{R}$ is unit matrix corresponding to right handed charged lepton singlet fields. 
In charged lepton basis the effective Majorana mass matrix is given by
\begin{equation}
M_{\nu}= 
  \begin{pmatrix}
    \frac{c}{9}+ \frac{x^{2}}{3M}& \frac{d}{9} & 0  \\
    \frac{d}{9} &  \frac{y^{2}}{3N} & \frac{c}{9}  \\
    0 & \frac{c}{9} & \frac{d}{9}
   \end{pmatrix}
\end{equation}
which symbolically can be written as
\begin{equation}
M_{\nu}= 
  \begin{pmatrix}
    X & \Delta  & 0  \\
    \Delta & X & X  \\
    0 & X & \Delta
   \end{pmatrix},
\end{equation}
where $\Delta$ denotes the equality between elements and X denotes arbitrary non-zero elements. In literature, such type of neutrino mass matrix structure is referred as hybrid textures\cite{ht1,ht2,ht3}. On changing the assignments of the fields we can have two more hybrid textures. For example, if we assign $\nu_{2}\sim \bf{1'}$, $\chi \sim \bf{1'}$ and $\Delta_{1}\sim \bf{1'}$ we end up with

\begin{equation}
\begin{pmatrix}
  X & X  & \Delta  \\
    X & \Delta & 0  \\
    \Delta & 0 & X
\end{pmatrix}.
\end{equation}

Similarly, the field assignments $\nu_{2}\sim \bf{1'}$, $\chi \sim \bf{1'}$ and $\Delta_{2}\sim \bf{1'}$ result in effective neutrino mass matrix   
  
\begin{equation}
\begin{pmatrix}
  X & 0  & \Delta  \\
    0 & \Delta & X  \\
    \Delta & X & X
\end{pmatrix}.
\end{equation}
In the next section, we study the phenomenological consequences of these neutrino mass matrices.

\section{Phenomenological consequences of the model }

In charged lepton basis, the effective Majorana neutrino mass matrix, $M_\nu$ is given by
\begin{equation}
 M_{\nu} = VM_{\nu}^{diag}V^{T},
\end{equation}
 where $V=U.P$ and 
$$ M_{\nu}^{diag} = {\begin{pmatrix}
    m_{1}& 0  & 0  \\
    0 & m_{2} & 0  \\
    0 & 0 & m_{3}
   \end{pmatrix}}. $$
   
   $U$ is Pontecorvo-Maki-Nakagawa-Sakata(PMNS) matrix and in standard PDG representation is given by

\begin{equation}
U={
  \begin{pmatrix}
   c_{12}c_{13}  & s_{12} c_{13} & s_{13}e^{-i\delta}  \\
    -s_{12}c_{23}-c_{12}s_{13}s_{23} e^{i\delta} & c_{12}c_{23}-s_{12}s_{13}s_{23}e^{i\delta}& c_{13}s_{23}  \\
    s_{12} s_{23}-c_{12}s_{13}c_{23}e^{i\delta} & -c_{12}s_{23}-s_{12}s_{13}c_{23}e^{i\delta}& c_{13}c_{23}
   \end{pmatrix}},
\end{equation}
where $s_{ij}= \sin \theta_{ij}$ and $c_{ij}= \cos \theta_{ij}$.
 The phase matrix, $P$ is 
 $$ {\begin{pmatrix}
    1& 0  & 0  \\
    0 & e^{2i\alpha} & 0  \\
    0 & 0 & e^{2i(\beta+\delta)}
   \end{pmatrix}}, $$
  where $\alpha$, $\beta$ are Majorana phases and $\delta$ is Dirac-type $CP$ violating phase.

  The neutrino mass model described by Eqn.(10) imposes two conditions on the neutrino mass matrix $M_\nu$, viz.:
\begin{eqnarray}
\nonumber
 &&\left(M_\nu\right)_{ab}=0,\\
 &&\left(M_\nu\right)_{uv}=\left(M_\nu\right)_{mn},
\end{eqnarray}
where $a=u=1$, $b=m=n=3$ and $v=2$ for neutrino mass matrix in Eqn.(9).

 It leads to two complex equations amongst nine parameters, viz.: three neutrino masses($m_1, m_2, m_3$), three  mixing angles($\theta_{12}, \theta_{23}, \theta_{13}$) and three $CP$ violating phases($\delta, \alpha, \beta$)
 \begin{equation}
  m_{1} U_{a 1}U_{b 1} +m_{2} U_{a 2}U_{b 2}e^{2i\alpha} +m_{3} U_{a 3}U_{b 3} e^{2i(\beta+\delta)}=0,
 \end{equation}
 and 
 \begin{eqnarray}
 \nonumber
 &&m_{1} (U_{u1}U_{v1}-U_{m1}U_{n1})+m_{2}(U_{u2}U_{v2}-U_{m2}U_{n2})e^{2i\alpha}\\
 &&  +m_{3}(U_{u3}U_{v3}-U_{m3}U_{n3})e^{2i(\beta+\delta)}=0.
 \end{eqnarray}
 
 \par We solve Eqn. (16) and (17) for mass ratios $\frac{m_1}{m_3}$ and $\frac{m_2}{m_3}$

 \begin{eqnarray}
  R_{13}\equiv \left|\frac{m_{1}}{m_{3}}e^{-2i(\beta+\delta)}\right|= \left|\frac{U_{a3}U_{b3}U_{u2}U_{v2}-U_{a2}U_{b2}U_{u3}U_{v3}+U_{a2}U_{b2}U_{m3}U_{n3}-
  U_{a3}U_{b3}U_{m2}U_{n2}}  {U_{a2}U_{b2}U_{u1}U_{v1}-U_{a1}U_{b1}U_{u2}U_{v2}+
  U_{a1}U_{b1}U_{m2}U_{n2}-U_{a2}U_{b2}U_{m1}U_{n1}}\right|,\\ 
  R_{23}\equiv\left|\frac{m_{2}}{m_{3}}e^{2i(\alpha-\beta-\delta)}\right|=\left|\frac{U_{a1}U_{b1}U_{u3}U_{v3}-U_{a3}U_{b3}U_{u1}U_{v1}+U_{a3}U_{b3}U_{m1}U_{n1}-
  U_{a1}U_{b1}U_{m3}U_{n3}}  {U_{a2}U_{b2}U_{u1}U_{v1}-U_{a1}U_{b1}U_{u2}U_{v2}+
  U_{a1}U_{b1}U_{m2}U_{n2}-U_{a2}U_{b2}U_{m1}U_{n1}}\right|,
 \end{eqnarray}
 where the ratios $R_{13}\equiv\frac{m_1}{m_3}$ and $R_{23}\equiv\frac{m_2}{m_3}$. The ratio $R_{23}$ can be obtained from $R_{13}$ using the transformation $\theta_{12}\rightarrow \frac{\pi}{2}-\theta_{12}$. The mass ratios $R_{13}$ and $R_{23}$ along with measured neutrino mass-squared differences provide two values of $m_3$, viz.: $m_3^a$ and $m_3^b$, respectively and is given by
  \begin{eqnarray}
  &&m_{3}^{a} = \sqrt{\frac{\Delta m^{2}_{12}+\Delta m^{2}_{23}}{1-R^{2}_{13}}},\\
  &&m_{3}^{b} = \sqrt{\frac{\Delta m^{2}_{23}}{1-R^{2}_{23}}}.
 \end{eqnarray}   
 
 These two values of $m_3$ must be consistent with each other, which results in
  \begin{equation}
  R_{\nu}=\frac{R_{23}^{2}-R_{13}^{2}}{|1-R_{23}^{2}|}\equiv\frac{\Delta m^{2}_{21}}{|\Delta m^{2}_{23}|}.
 \end{equation}
The ratios $\frac{m_1}{m_3}e^{-2i(\beta+\delta)}$ and $\frac{m_2}{m_3}e^{2i(\alpha-\beta-\delta)}$, to first order in $s_{13}$, is given by
\begin{eqnarray}
\nonumber
&&\frac{m_1}{m_3}e^{-2i(\beta+\delta)}\approx-\frac{c_{23}^2}{s_{23}^2 }+\frac{e^{-i\delta}s_{13} \left(s_{23}^2+c_{23}^2 e^{2i\delta}\right) \left(s_{12}-c_{12}c_{23} s_{23}^2\right)}{s_{12} s_{23}^5},\\ \nonumber
&&\frac{m_2}{m_3}e^{2i(\alpha-\beta-\delta)}\approx-\frac{c_{23}^2}{s_{23}^2 }+\frac{e^{-i\delta}s_{13} \left(s_{23}^2+c_{23}^2 e^{2i\delta}\right) \left(c_{12}+s_{12}c_{23} s_{23}^2\right)}{c_{12} s_{23}^5}.\\
\end{eqnarray}
Using these approximated mass ratios we find
 \begin{eqnarray}
R_{13}^2\approx\frac{c_{23}^4}{s_{23}^4}-\frac{s_{13}\left(s_{12}-c_{12}c_{23}s_{23}^2\right)}{s_{12}s_{23}^7}\left(2c_{23}^2\cos\delta-\frac{s_{13}}{s_{12}s_{23}^3} \left(s_{12}-c_{12}c_{23}s_{23}^2\right)\left(c_{23}^4+s_{23}^4+2c_{23}^2s_{23}^2\cos2\delta\right)\right), \\
R_{23}^2\approx\frac{c_{23}^4}{s_{23}^4}-\frac{s_{13}\left(c_{12}+s_{12}c_{23}s_{23}^2\right)}{c_{12}s_{23}^7}\left(2c_{23}^2\cos\delta-\frac{s_{13}}{c_{12}s_{23}^3} \left(c_{12}+s_{12}c_{23}s_{23}^2\right)\left(c_{23}^4+s_{23}^4+2c_{23}^2s_{23}^2\cos2\delta\right)\right),
 \end{eqnarray}
 and
 \begin{equation}
R_{23}^2-R_{13}^2\approx s_{13}\frac{\left(4\sin2\theta_{12}-\sin 2 \theta_{23}\right)\left(c_{23}^4+s_{23}^4+2c_{23}^2s_{23}^2\cos 2\delta\right)s_{13}-4\sin 2\theta_{12}\sin^3 2\theta_{23}\cos\delta}{8\sin 2\theta_{12}s_{23}^8}.
\end{equation}
$m_2$ must be greater than $m_1$, or equivalently, $R_{23}^2-R_{13}^2>0$ which is possible if
 
 \begin{eqnarray}
(4\sin2\theta_{12}-\sin 2 \theta_{23})(c_{23}^4+s_{23}^4+2c_{23}^2s_{23}^2\cos 2\delta)s_{13}>4\sin 2\theta_{12}\sin^3 2\theta_{23}\cos\delta),
 \end{eqnarray}
 which translates to constraint on $\delta$ given by
 
 \begin{eqnarray}
 \left(\frac{4\sin2\theta_{12}\sin^3 2 \theta_{23}}{\left(4\sin2\theta_{12}-\sin2\theta_{23}\right)\left(c_{23}^4+s_{23}^4\right)s_{13}}\right)\cos\delta -
 \left(\frac{\sin^2 2\theta_{23}}{2\left(c_{23}^4+s_{23}^4\right)}\right)\cos2\delta <1.
 \end{eqnarray}
 Using the experimental data shown in Table \ref{tab2}, we find that $\delta$ can take values only near $\delta\approx 90^o$ or $\delta\approx 270^o$ for the model to be  consistent with solar mass hierarchy. However, it will, further, get constrained by the requirement of $R_\nu$ to be within its experimental range. For normal hierarchy(NH), $R_{23}^2-R_{13}^2>0$ and $1-R_{23}^2>0$(or $\frac{R_{13}}{R_{23}}<1$ and $R_{23}<1$). Substituting $\delta\approx90^o$ or $270^o$ in Eqn. (25), the condition $1-R_{23}^2>0$ (to leading order in $s_{13}$) yields $\cot^4\theta_{23}<1$ i.e. $\theta_{23}$ is above maximality($\theta_{23}>45^o$). Similarly, for inverted hierarchy(IH), the condition $R_{23}^2-R_{13}^2>0$ and $1-R_{13}^2<0$(or $\frac{R_{13}}{R_{23}}<1$ and $R_{13}>1$) predicts $\theta_{23}$ below maximality($\theta_{23}<45^o$).

With the help of constraints derived  for $\delta$, $\theta_{23}$ for NH as well as IH, it is straightforward to show that $R_{\nu}$(Eqn.(22)) is $\mathcal{O}(10^{-2})$ for NH and  $\mathcal{O}(10^{-1})$ for IH. For example, if $\delta=87.6^o$, $\theta_{23}=48^o$, $\theta_{12}=34.5^o$ and $\theta_{13}=8.5^o$, $R_{\nu}$ is 0.025 for NH. Similarly, if $\delta=87.6^o$, $\theta_{23}=44^o$, $\theta_{12}=34.5^o$ and $\theta_{13}=8.5^o$, $R_{\nu}$ is 0.849 for IH. Thus, the requirement that model-prediction for $R_{\nu}$ must lie within its experimentally allowed range hints towards normal mass hierarchy.  These approximated analytical results will be extremely helpful to comprehend the phenomenological predictions obtained from the numerical analysis which is based on the exact constraining Eqns.(18) and (19).  
 
In numerical analysis, we have used Eqn.(22) as our constraining equation to obtain the allowed parameter space of the model i.e. $R_{\nu}$ must lie within its 3$\sigma$ experimental range. We have used latest global-fit data shown in Table \ref{tab2}.
 \begin{table}
\centering
\begin{tabular}{llc}
\hline
Parameters  &  Best-fit$\pm 1\sigma$ & 3$\sigma$ range\\
\hline
 $\Delta m_{21}^2[10^{-5}$ eV$^2$] & $7.55^{+0.20}_{-0.16}$ & $7.05-8.14$ \\
 $\Delta m_{31}^2[10^{-3}$ eV$^2$] (NH) & $2.50\pm0.03$ & $2.41-2.60$ \\
 $\Delta m_{31}^2[10^{-3}$ eV$^2$] (IH)& $2.42^{+0.03}_{-0.04}$ & $2.31-2.51$ \\
 Sin$^2 \theta_{12}/10^{-1}$ & $3.20^{+0.20}_{-0.16}$ & $2.73-3.79$ \\
 Sin$^2 \theta_{23}/10^{-1}$ (NH) & $5.47^{+0.20}_{-0.30}$ & $4.45-5.99$ \\
 Sin$^2 \theta_{23}/10^{-1}$ (IH) & $5.51^{+0.18}_{-0.30}$ & $4.53-5.98$ \\
 Sin$^2 \theta_{13}/10^{-2}$ (NH) & $2.160^{+0.083}_{-0.069}$ & $1.96-2.41$ \\
 Sin$^2 \theta_{13}/10^{-2}$ (IH) & $2.220^{+0.074}_{-0.076}$ & $1.99-2.44$ \\
\hline
\end{tabular}
\caption{\label{tab2}The latest global-fit results of neutrino mixing angles and neutrino mass-squared differences used in this analysis \cite{data}.}
\end{table} 
The experimentally known parameters such as mass-squared differences and mixing angles are randomly generated with Gaussian distribution whereas 
$CP$ violating phase $\delta$ is allowed to vary in full range ($0^o-360^o$) with uniform distribution($\approx10^{7}$ points). The mass ratios $R_{13}$ and $R_{23}$ depend on $\theta_{12},\theta_{23},\theta_{13}$ and $\delta$. Using experimental data shown in Table \ref{tab2}, we first calculate the prediction of the model for $R_{\nu}$ with normal as well as inverted hierarchy. It is evident from Fig. 1 that $R_{\nu}$ is $\mathcal{O}(10^{-1})$ for IH i.e. outside the experimental 3$\sigma$ range of $R_\nu$ which is, also, in consonance with above analytical discussion. Hence, inverted hierarchy (IH) is ruled out at more than 3$\sigma$. 

\begin{figure}
\centering
\includegraphics[scale=0.7]{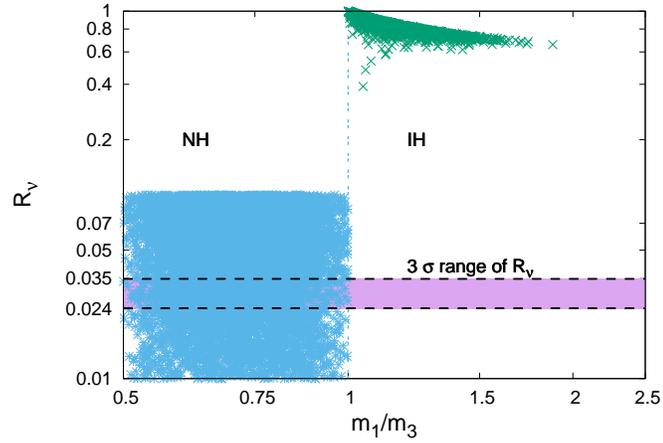}
\caption{\label{fg1} $R_{\nu}$ as a function of $m_{1}$/$m_{3}$ for normal hierarchy (NH) and inverted hierarchy (IH).}
\end{figure}
In Fig. 2(a), we have depicted correlation between $\delta$ and $\theta_{23}$ at 3$\sigma$. $\theta_{23}=45^o$ is not allowed because $1-R_{23}^2$ must be less than 1. Also, the point ($\theta_{23}=45^o, \delta=90^o$ or $270^o$) is not allowed otherwise $R_{\nu}<0$. $\theta_{23}$ is found to be above maximality and Dirac-type $CP$ violating phase $\delta$ is constrained to a very narrow region in I$^{st}$ and IV$^{th}$ quadrant. In Fig. 2(b), 2(c) and 2(d), we have shown the normalized probability distributions of $\theta_{23}$ and $\delta$. The 3$\sigma$ ranges of these parameters are given in Table \ref{tab4}. 

\begin{figure}
 \begin{center}
 \epsfig{file=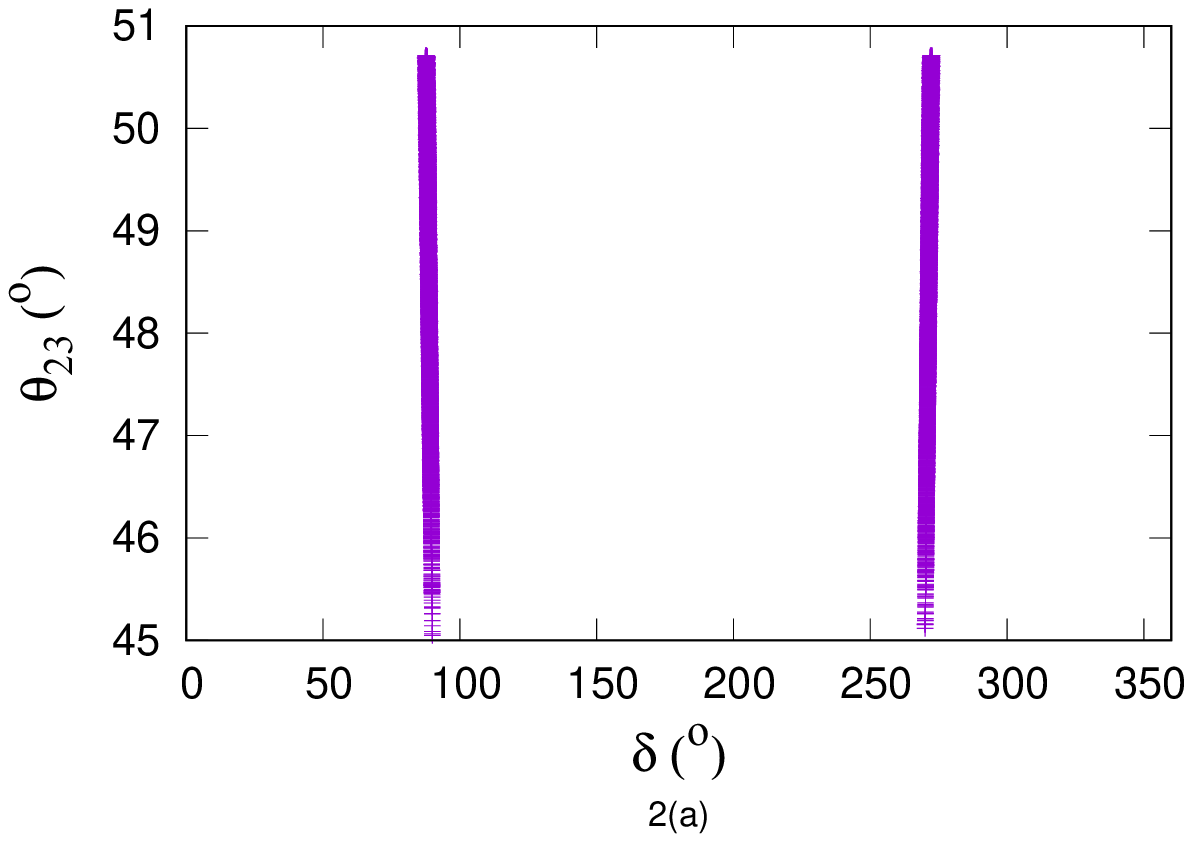,height=5.0cm,width=8.0cm}\\
{\epsfig{file=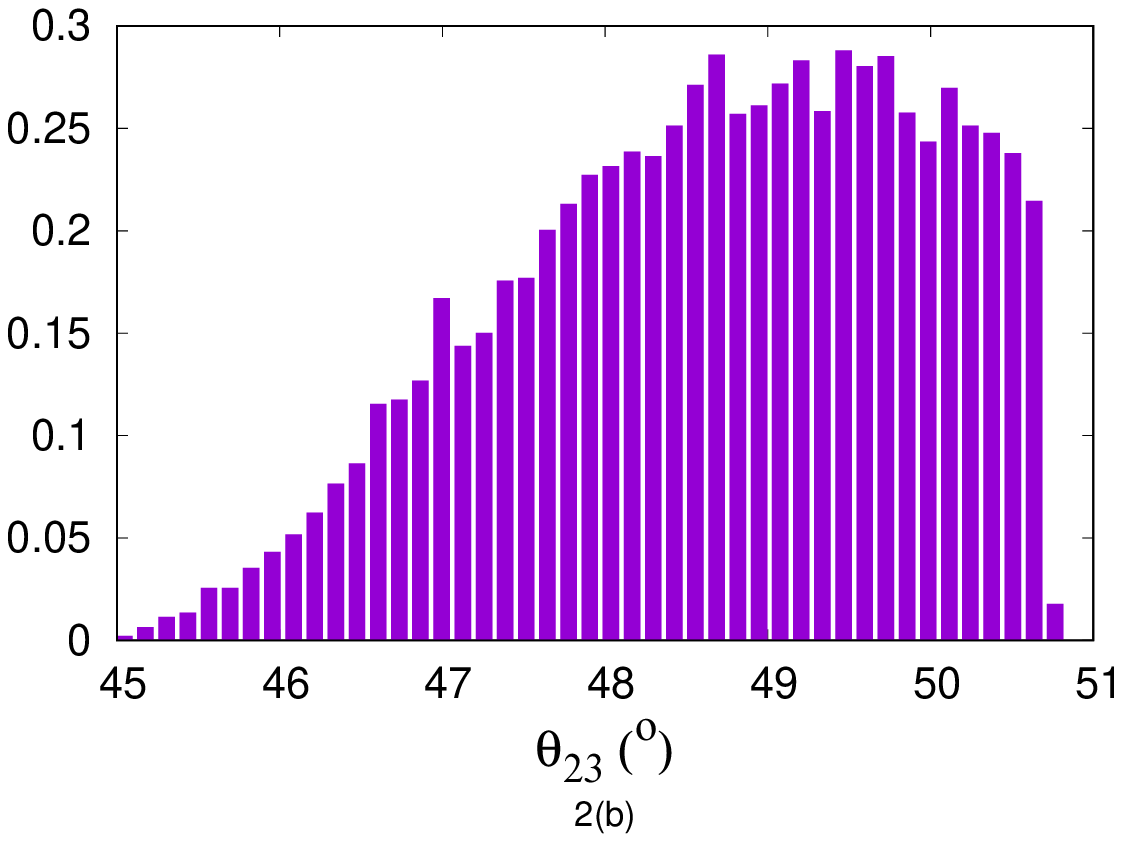,height=5.0cm,width=7.0cm},
\epsfig{file=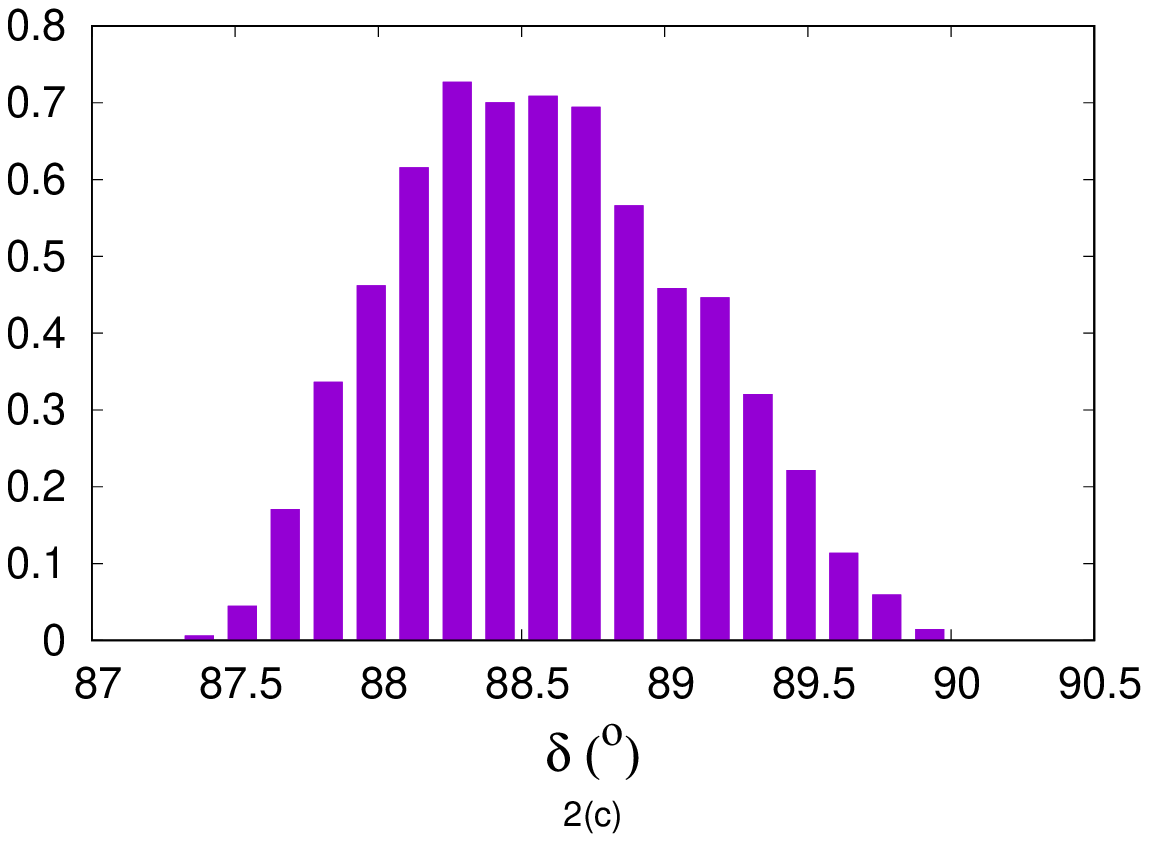,height=5.0cm,width=7.0cm},
\epsfig{file=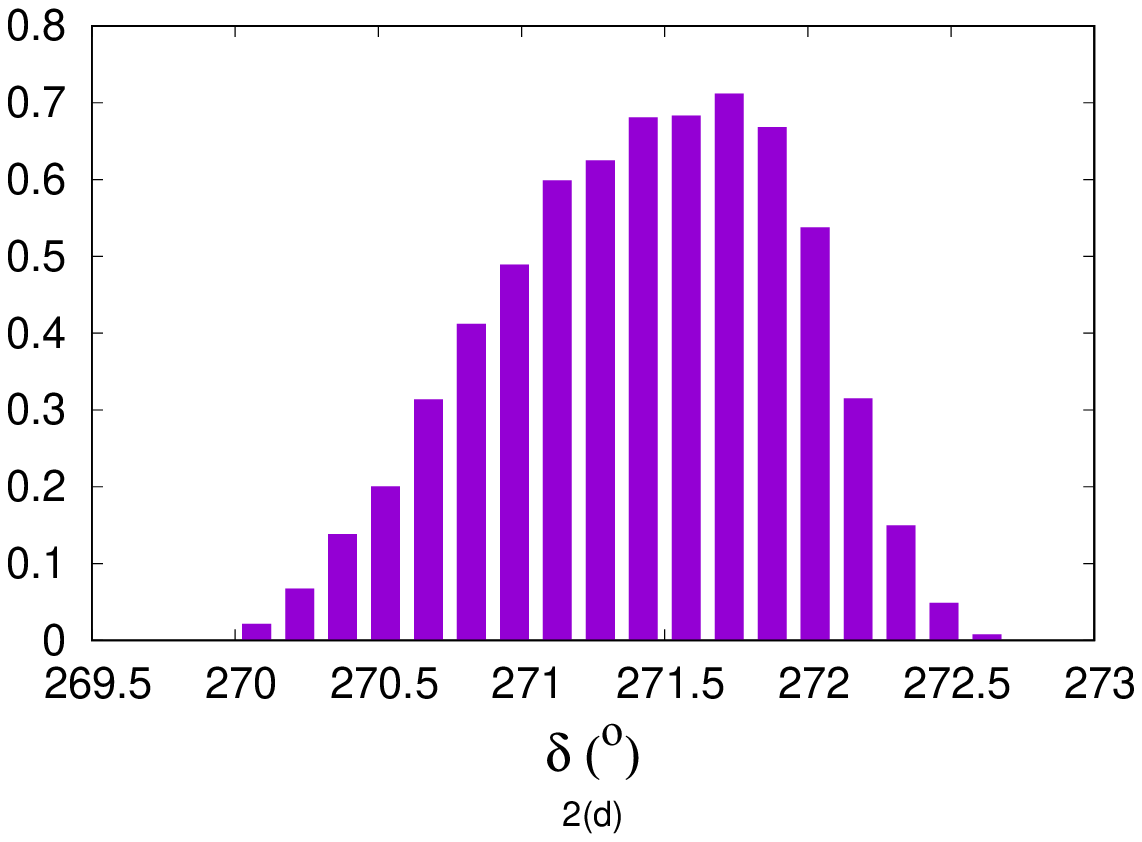,height=5.0cm,width=8.0cm}}
\end{center}
  \caption{\label{fig2} $\delta-\theta_{23}$ correlation plot at 3$\sigma$(Fig. 2(a)) and probability distribution plots for $\theta_{23}$ (Fig. 2(b)) 
  and $\delta$ (Fig. 2(c) and 2(d)).}
\end{figure}

\begin{figure}[H]
\centering
\includegraphics[scale=0.7]{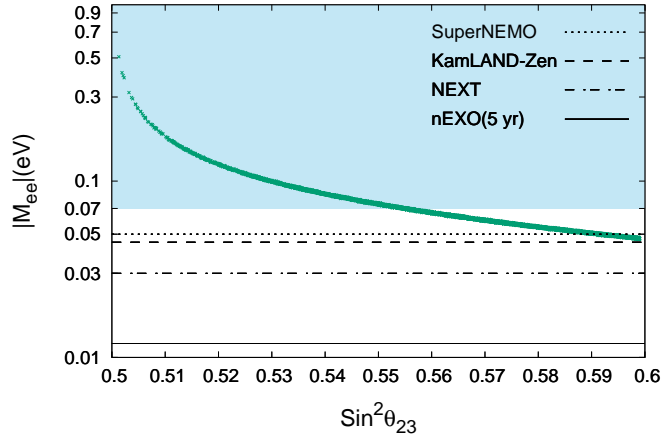}
\caption{\label{fg3} $sin^{2}\theta_{23}-|M_{ee}|$ correlation plot at 3$\sigma$. The current sensitivity of SuperNEMO, KamLAND-Zen, NEXT, nEXO(5 yr) for $|M_{ee}|$ is, also, shown. The shaded region depicts the $|M_{ee}|$-$\sin^2\theta_{23}$ parameter space disallowed by incorporating the cosmological bound on sum of neutrino masses i.e. $\sum_{i}m_i<0.24$ eV (at $95\%$ C.L.)\cite{cosmo} in our numerical analysis.}
\end{figure}

One of the desirable feature of a neutrino mass model is its prediction of the observable(s) which can be probed outside the neutrino sector. One such process is $0\nu\beta\beta$ decay, the amplitude of which is proportional to effective Majorana neutrino mass $|M_{ee}|$ given by 
\begin{equation} 
 |M_{ee}| = |m_{1} c_{12}^{2} c_{13}^{2} +m_{2} s_{12}^{2} c_{13}^{2} e^{2i\alpha} +m_{3} s_{13}^{2} e^{2i\beta}|.
\end{equation}

In Fig. 3, we have shown $sin^{2}\theta_{23}-|M_{ee}|$ correlation plot at 3$\sigma$. The important feature of the present model is the existence of lower bound $|M_{ee}|>0.047$eV(at 3$\sigma$) which is within the sensitivity reach of 0$\nu\beta\beta$ decay experiments like SuperNEMO \cite{nemo}, KamLAND-Zen \cite{zen}, NEXT \cite{next1,next2},
 nEXO\cite{nexo}. The $|M_{ee}|$-$\sin^2\theta_{23}$ parameter space is, further, constrained with inclusion of the cosmological bound on sum of neutrino masses($\sum_{i}m_i<0.24$ eV at 95$\%$ C.L., TT, TE, EE+lowE+lensing)\cite{cosmo} in our numerical analysis. In particular, there exist an upper(lower) bound on $M_{ee}\leq 0.070$ eV ($\theta_{23}\geq 48^o$) at 3$\sigma$. 
 
A similar analysis of neutrino mass matrices shown in Eqns.(11) and (12) reveals that these textures are not compatible with present global-fit data on neutrino masses and mixing including latest hints of normal hierarchical neutrino masses, higher octant of $\theta_{23}$ and near maximal Dirac-type $CP$ violating phase $\delta$\cite{data,newdata,t130,newdata2}. It has been already  shown in reference \cite{ht4} that hybrid textures are stable against the one loop RG effects. Consequent to the small renormalization group (RG) effects, the phenomenological consequences of hybrid texture  structure predicted at higher energy scale can be studied with same structure at electroweak scale.  Furthermore,  being a minimal model, it will have interesting implications for leptogenesis which will be discussed elsewhere.

The  extension of the scalar field sector with scalar singlet($\chi$) and scalar triplets($\Delta_{i}$) in addition to Higgs doublet$(\Phi_{i})$ may provide interesting phenomenology in collider experiments. The important feature of Higgs triplet model is presence of doubly charged scalar boson($\Delta^{\pm \pm}$) in addition to $H^{\pm}W^{\mp}Z$ vertex at tree level\cite{collider}. The decay of doubly charged bosons($\Delta^{\pm \pm}$) to charged leptons connects the neutrino sector and collider physics of the model. Due to involvement of the same couplings in neutrino sector and decay of $\Delta^{\pm \pm}$, collider phenomenology can be studied independently. With two scalar triplets $(\Delta_{1,2})$, the physical Higgs are related to doubly charged scalar bosons$(\Delta^{++})$ as
$$
\begin{pmatrix} 
H_{1}^{++}  \\
H_{2}^{++}
\end{pmatrix}
=
\begin{pmatrix}
\cos\theta & \sin\theta \\
-\sin\theta & \cos\theta 
\end{pmatrix}
\quad
\begin{pmatrix} 
\Delta_{1}^{++}  \\
\Delta_{2}^{++}
\end{pmatrix}
$$
where $\theta$ is mixing angle. These physical Higgs can decay through different decay channels such as $H^{++}\rightarrow l^{+}l^{+},H^{++}\rightarrow W^{+}W^{+},H^{++}\rightarrow H^{+}W^{+},H^{++}\rightarrow H^{+}H^{+}$. The last two decay channels might be kinematically suppressed as they depend on the mass difference of $H^+$ and $H^{++}$. In particular, the decay mode $H^{++}\rightarrow l^{+}l^{+}$ is dominant, however, it will depend on the VEV acquired by the scalar triplets\cite{mitra}. For this decay mode, branching ratio depends on Yukawa couplings i.e. structure of neutrino mass matrix. The observation of dilepton decay mode, in collider experiment may, in general, shed light on the neutrino mass hierarchy.

 \begin{table}
\centering
\begin{tabular}{ccc}
\hline
                                          $\theta_{23}$  & $\delta$ & $|M_{ee}|$  \\ 
                                                 (bfp, 3$\sigma$ range) & (bfp(s), 3$\sigma$ range(s))&  (3$\sigma$ lower bound) \\  
\hline
                                     $48.68 ^{o}$                        & $88.58^{o}$ & $\geq$0.047 eV               \\
                                        $45.17^{o}-50.70^{o}$                & $87.50^{o}-89.70^{o}$ &                \\
                                                                          & $271.42^{o}$ &                          \\
                                                                          & $270.20^{o}-272.50^{o}$ &\\
																																					\hline                                                                    
\end{tabular}

\caption{ \label{tab4} Prediction of the model for $\theta_{23}$, $\delta$ and $|M_{ee}|$.}
\end{table}

\section{Conclusions}
\par In conclusion, we have presented a neutrino mass model based on $A_{4}$ flavor symmetry for leptons within type-I+II seesaw framework. The model is economical in terms of extended scalar sector and is highly predictive. The field content assumed in this work predicts three textures for $M_{\nu}$ based on the charge assignments under $SU(2)_L$ and $A_4$. However, only one(Eqn.(10)) is found to be compatible with experimental data on neutrino masses and mixing angles. We have studied the phenomenological implications of this texture in detail. The solar mass hierarchy i.e. $R_{23}^2-R_{13}^2>0$ constrains Dirac-type $CP$ violating $\delta$ to narrow ranges $87.50^{o}-89.70^{o}$ and $270.20^{o}-272.50^{o}$ at 3$\sigma$. The sharp correlation between Dirac-type $CP$ violating phase $\delta$ and atmospheric mixing angle $\theta_{23}$ demonstrates the true predictive power of the model(Fig. 2(a)). The predictions for these less precisely known oscillation parameters($\delta$ and $\theta_{23}$) are remarkable which can be tested in neutrino oscillation experiments like T2K, NO$\nu$A, SK and DUNE to name a few. We have, also, calculated effective Majorana neutrino mass $|M_{ee}|$. The important feature of the model is existence of lower bound on $|M_{ee}|$ which can be probed in $0\nu\beta\beta$ decay experiments like SuperNEMO, KamLAND-Zen, NEXT and nEXO. The main predictions of the model are:
\begin{enumerate}[label=\roman{*}.]
\item normal hierarchical neutrino masses.
\item $\theta_{23}$ above maximality($\theta_{23}>45^o$).
\item near-maximal Dirac-type $CP$ violating phase $\delta = 88.58^o$ or $271.42^o$.
\item $3\sigma$ range of effective Majorana neutrino mass $0.047\leq |M_{ee}|\leq 0.070$ eV.
\end{enumerate}
A precise measurements of Dirac-type $CP$ violating phase $\delta$, neutrino mass hierarchy and $\theta_{23}$ is important to confirm the viability of the model presented in this work. 

\vspace{1cm}
\textbf{\Large{Acknowledgments}}\\
The authors thank R. R. Gautam for useful discussions. S. V. acknowledges the financial support provided by UGC-BSR and DST, Government of India vide Grant Nos. F.20-2(03)/2013(BSR) and MTR/2019/000799/MS, respectively. M. K. acknowledges the financial support provided by Department of Science and Technology, Government of India vide Grant No. DST/INSPIRE Fellowship/2018/IF180327. The authors, also, acknowledge Department of Physics and Astronomical Science for providing necessary facility to carry out this work.

\end{document}